# Improving the long-term stability of new-generation perovskite-based TCO using binary and ternary oxides capping layers

Moussa Mezhoud[1*], Martando Rath[1], Stéphanie Gascoin[1], Sylvain Duprey[2], Philippe Marie[2], Julien Cardin[2], Christophe Labbé[2], Wilfrid Prellier[1], and Ulrike Lüders[1*]

[1]*CRISMAT, CNRS, Normandie Univ, ENSICAEN, UNICAEN, 6, boulevard du Maréchal Juin, F-14050 Caen, France.*
[2]*CIMAP, CNRS, ENSICAEN, UNICAEN, CEA, Normandie Univ, 6, boulevard du Maréchal Juin, F-14050 Caen, France.*

*[*moussa.mezhoud@ensicaen.fr](mailto:moussa.mezhoud@ensicaen.fr) ; [ulrike.luders@ensicaen.fr](mailto:ulrike.luders@ensicaen.fr)*

## ABSTRACT

We report the impact of capping layers on vanadate based transparent conductive oxides (TCOs) to prolong the thermal stability with a minimal loss of resistivity during heat treatment in ambient environment. In the present study, various protecting layers (amorphous $Al_2O_3$, $LaAlO_3$ (LAO), $TiO_2$ grown in base pressure and $TiO_2$ deposited under oxygen partial pressure) are grown *in-situ* on polycrystalline perovskite $SrVO_3$ (SVO) thin films using Pulsed Laser Deposition (PLD). The results show that amorphous $LaAlO_3$ is the most promising protection layer among the oxide layers, to preserve both electrical and optical properties of perovskite SVO films from natural as well as artificial aging. Our present approach for a capping layer on SVO may address the long-term stability issues of correlated TCOs and would open an opportunity for the future oxide electronics applications.

## INTRODUCTION

The progress in creating next-generation optoelectronic devices such as flat-panel displays, smartphones, sensors, and solar cells is intricately tied to the developpement of a vital component in these devices: transparent and conducting oxides (TCOs)[1–7]. These materials possess a unique combination of electrical conductivity and optical transparency, two competing properties that are notoriously challenging to achieve within a single material. Recently, a new family of transparent and conducting materials has been discovered in the oxydes perovskite of structure $ABO_3$[8–10], arousing great interest in the scientific community.

$SrVO_3$ (SVO) emerges as a highly promising material[8,11–15], offering a compelling alternative to commonly used transparent conducting oxides like Indium Tin Oxide (ITO)[16], due to its remarkable optical and electrical properties, along with the abundant and low-cost nature of its chemical elements constituent. However, numerous studies have pointed to the critical issue of SVO instability when exposed to oxygen-containing atmospheres[17–20]. The degradation, primarily attributed to the presence of $V^{4+}$, an unstable oxidation state[21], has been observed to lead to the formation of insulating over-oxidized phases $V^{5+}$ phases, such as $Sr_2V_2O_7$ or Sr-rich $Sr_3V_2O_8$ on the surface of the SVO film[22]. Due to their electrically insulating nature, these phases drastically alter the surface conductivity and are known to be soluble in water[23–25], which provides a significant barrier to the long-term use of SVO as a TCO. Moreover, the surface instability of perovskite oxides is a well known issue when exposed to different environmental conditions, such as variations in temperature, humidity, and gas atmospheres, more specifically, which contains oxygen. This phenomenon gives rise to surface reconstruction and segregation, formation of surface defects, and phase transformations, all of which can drastically alter the material's properties[26–29]. Consequently, surface instability has emerged as a pivotal factor that impacts the functionality, stability, and durability of perovskite oxide-based devices and technologies. These drawbacks of the vanadate TCOs may provide an important obstacle to widespread industrial acceptance of this extremely promising new TCO.

In order to address this issue, the concept of using capping layers to engineer and enhance the properties of perovskite oxides has gained significant attention in recent years[30–34]. Capping layers, which are very thin films deposited on the surface of perovskite oxide materials, offer a powerful approach to tailor the structural, electrical, and chemical characteristics of these functional materials. For example, an enhacement in the Curie temperature of $SrRuO_3$ thin films has been reported using a capping layer of $SrTiO_3$ (STO) that reduces the oxygen octahedral rotation[35]. STO has also been utilized to stabilize the $Sr_3Al_2O_6$ sacrificial layer in ambient atmosphere and for growth of functional oxides at high temperatures[36,37].

For vanadates-based perovskite oxides, capping layers has been used to stabilize the unstable $V^{3+}$ and $V^{4+}$ oxidation states. Kumar *et al.*[38] employed a thin film of 8 nm $LaAlO_3$ as the capping layer of $PrVO_3$ (PVO) films to prevent oxydation of $V^{3+}$ to $V^{4+}$, which has considerably reduced the magentic dead layer in PVO films. Recently, Caspi *et al.*[39] stabilized the V[4+] oxidation state of the surface of epitaxial SVO thin films by using an amorphous TiO$_2$ capping layer, however, in this work the authors have only studied the stability at ambient conditions. In an earlier study, we have investigated the accelerated aging process of SVO films at different temperatures (150°C to 250°C) and observed that when exposed to 250°C, SVO films without capping layers lose their crystalline structure and become insulating[21], so in the present study we investigate the effect of using different capping layers reported in litterature such as TiO$_2$, LAO, and Al$_2$O$_3$ on the stability of SVO films at 250°C, the temperature at which unprotected films completely degrade. The intention of exposing the material to non-ambient conditions like an elevated temperature is to accelerate the aging process and provide a predicition of the materials's stability with different capping layers over an extended period, thereby identifying the most effective capping layer.

## RESULTS AND DISCUSSION

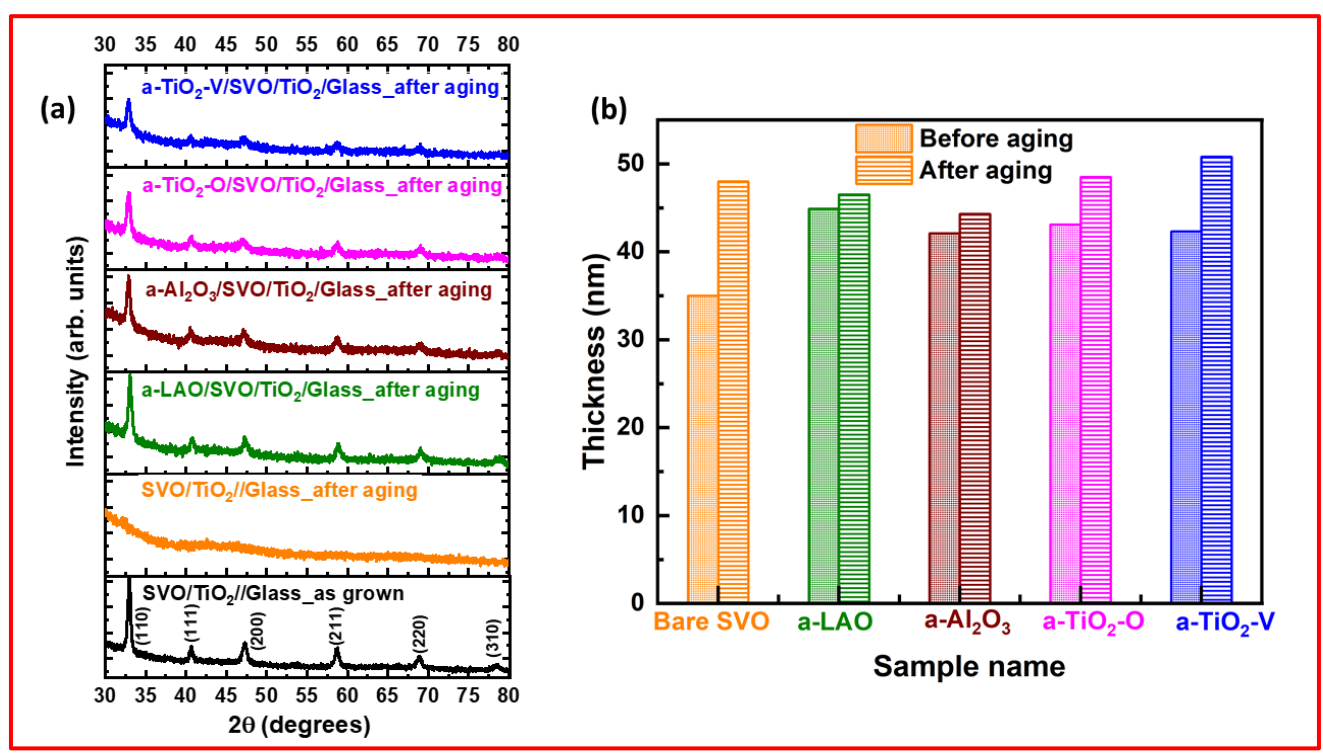

**Figure 1.** (a) GIXRD patterns of SVO thin films grown on TiO$_2$ buffered glass substrate with different protection layers (PL) and without PL, after heat treatment at 250 ºC for 24h in air. (b) Effect of annealing temperature on the thickness of the samples before and after heat treatment.

Figure 1.a depicts the GIXRD patterns of SVO thin films grown on TiO$_2$ buffered glass substrate without protection layer and with four different protection layers (PL) named a-LAO, a-Al$_2$O$_3$, a-TiO$_2$-V and a-TiO$_2$-O, after annealing at 250 ºC for 24h. For the SVO film without capping layer (orange line), all the diffraction peaks related to the SVO phase (Pm-3m) have vanished and the sample exhibits an amorphous nature as observed previously[21]. For the films with capping layers, the presence of diffraction peaks of different crystallographic plans such as (110), (111), (200), (211), (220) and (310) of the SVO phase confirms the polycrystalline nature of the films, even after an annealing at 250 ºC for 24h. Furthermore, we have not found any peaks related to secondary phases in the XRD patterns, confirming the stability of the SVO phase after 24h of heat treatment in the limits of the technique. Among all capping layers, SVO films with a-LAO, a- $Al_2O_3$ and a-$TiO_2$-O show clear and distinguished peaks even at higher 2θ angles. However, the sample with a-$TiO_2$-V capping layer shows slightly smaller intensities compared with the other samples, suggesting a lower protection effect of the a-$TiO_2$-V capping layers.

Figure 1.b shows the role of annealing temperature on the evolution of the thickness of the SVO samples before and after aging. The film thickness of non-capped SVO films is known to increase during an annealing at 250°C for 24 hours[21], which was attributed to the introduction of oxygen atoms into the films. Here, the non-capped SVO film annealed at 250°C (orange color) shows a similar significant thickness increase (from 35 nm to 48 nm, *i.e.* 27%), which is not observed for the capped films. The thickness of SVO with a-LAO and a- $Al_2O_3$ protection layers have a minimal increase *i.e.* around 3 and 5 %, respectively, under heat treatment at 250 ºC for 24h in air. However, the thickness of the sample with a-$TiO_2$ layers deposited under oxygen atmosphere or in vacuum, shows a stronger thickness increase (*i.e.* 11 and 17 %, respectively, compared to the as grown film). The GIXRD and XRR measurements show that after heat treatment at 250°C, SVO films with a-LAO and a- $Al_2O_3$ protection layers preserve their crystalline structure without further increase of thickness. However, for SVO films protected with a-$TiO_2$ capping layers, the intensity of diffraction peaks diminishes with increasing the total thickness of the films. These two observations point to a weaker protective character of the $TiO_2$ layers, leading to a partial degradation of SVO films.

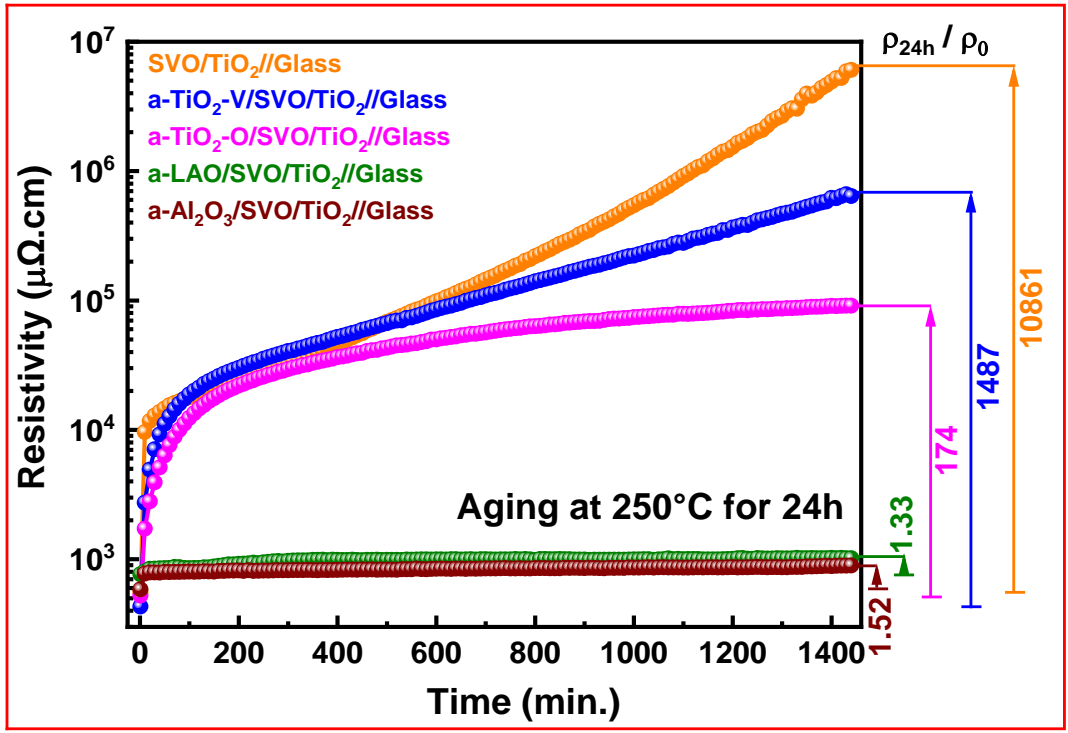

**Figure 2.** The variation of electrical resistivity of SVO thin films with and without protection layers (PLs), during heat treatment of the samples on a hot plate at 250 ºC for 24h in air atmosphere.

Figure 2 shows the evolution of electrical resistivity of the polycrystalline SVO films with and without protection layers during the heat treatment. As expected[21], the SVO film without capping layer shows an increasing resistivity. After 24h, the sample turns into a bad conductor with the destruction of its perovskite structure, which is confirmed from GIXRD results. For SVO films with different capping layers, the SVO films with a-LAO and a-$Al_2O_3$ capping layers show during the heat treatments a small increase of the resistivity, quantified here by the ratio between the resistivity after 24h of aging ($\rho_{24h}$) and the initial resistivity value before aging ($\rho_0$) of 1.33 and 1.52, respectively. The samples with a-$TiO_2$-O and a-$TiO_2$-V capping layers show an increase of the resistivity until 180 minutes, after which the resistivity almost saturates for the sample with a-$TiO_2$-O capping layer. In contrast, the resistivity continues to increase for the sample with a-$TiO_2$-V capping layer. This sample shows an evolution with time similar to the unprotected SVO, but with a reduced resistivity value. The ratio $\rho_{24h}/\rho_0$ of SVO films with $TiO_2$ deposited in vacuum and under oxygen atmosphere is found to be 1487 and 174, respectively. These observations underline that protective effect is not the same for the different types of capping layers.

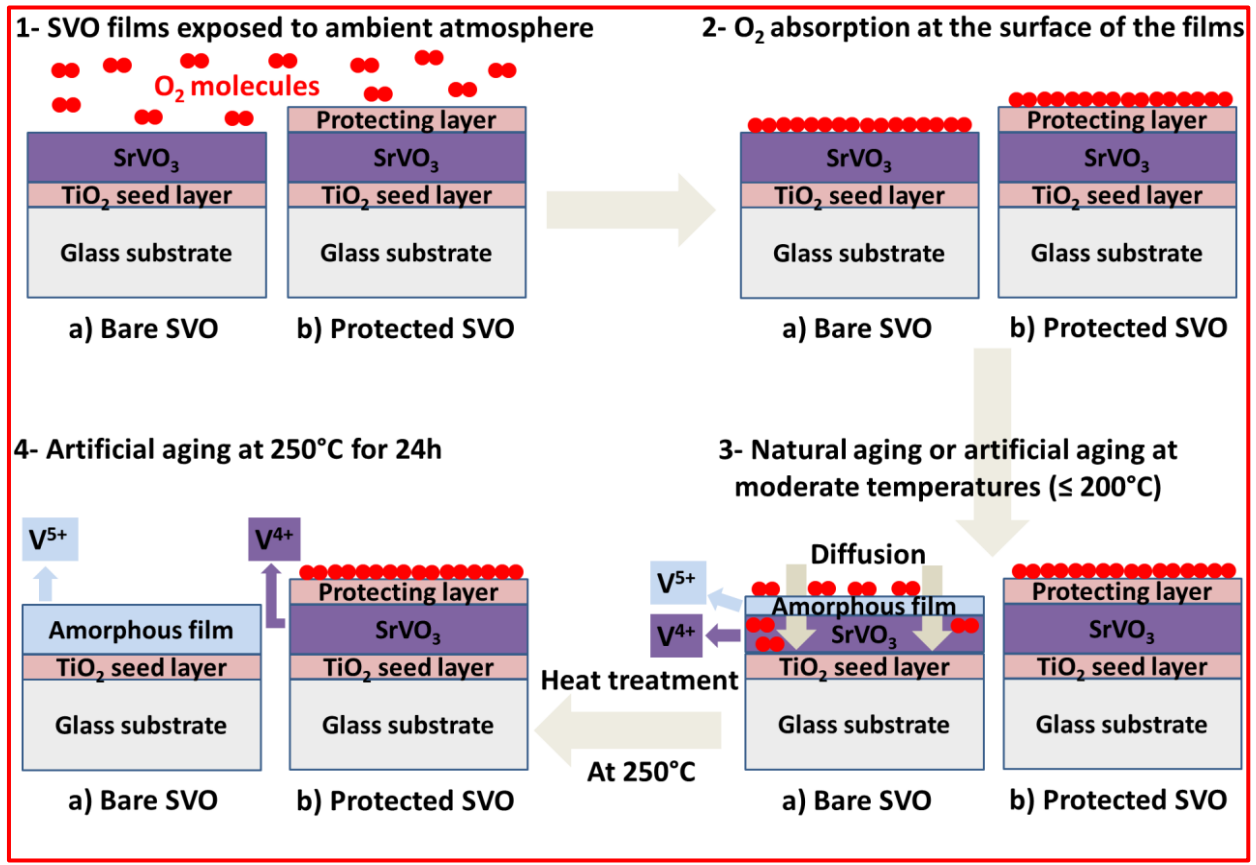

**Figure 3.** Illustration of the degradation process of a) bare SVO and b) protected SVO films in ambient atmosphere at moderate and high aging temperatures.

The substantial variation of resistivity of SVO films protected with $TiO_2$-based capping layers, compared to the minor variation observed using Al-based capping layers, is an indication that the $TiO_2$ films have a high permeability of O atoms[40,41]. $TiO_2$ protecting layers, regardless of whether they are deposited with or without oxygen, do not act as an anti-diffusion barrier for oxygen. Consequently, oxygen atoms present in the surrounding air are likely to diffuse into the SVO film through the $TiO_2$ protecting film like in the unprotected films with a slowed kinetics, and this process (illustrated in Figure 3) is further facilitated by elevated aging temperatures. In addition, it is known that $TiO_2$ possesses a higher oxygen diffusion coefficient ranging from four to nine times greater than that of $Al_2O_3$[42]. This property makes $TiO_2$ a popular choice as an active layer in ReRAM memories and memristors where a high oxygen diffusion rate is needed. However, in case of a-LAO and a-$Al_2O_3$ layers, the interaction of molecular oxygen with the Al ions forms a thermally stable surface on top of SVO. This passivating layer stops further diffusion of oxygen atoms to reach the underlying SVO layer (as illustrated in figure 3). Furthermore, $Al_2O_3$ is widely used as an oxygen anti-diffusion barrier in organic optoelectronic devices such as organic light emitting diodes or solar cells, especially when the polymers used are extremely sensitive to oxygen[43,44]. This characteristic of $Al_2O_3$ is clearly consistent with the findings provided in this study.

Figure 4 displays the surface topography of the SVO films before and after aging with various capping layers. We used the a-$TiO_2$-V capped film as a reference before aging. Prior to aging, the film exhibited a smooth surface with low RMS roughness, approximately 1 nm, over a large scan area of 10 µm x 10 µm, demonstrating the homogeneity of the as-grown film. Following aging, the film surface underwent significant alterations, characterized by the emergence of porosity and a substantial increase in RMS roughness, reaching 6 nm. Additionally, analysis of a smaller scan area of 500 nm x 500 nm revealed an observable increase in grain size compared to the as-grown film, indicative of microstructural changes induced by aging. These observations can be probably attributed to the absorption of atmospheric oxygen by the vacuum-deposited $TiO_2$ film, consistent with the observed increase in thickness after aging. For the SVO film protected with a-$TiO_2$-O capping layer, an increase in roughness and grain size is noticeable but remains less significant than that of the film protected with a-$TiO_2$-V layer, which highlights the impact of the $TiO_2$ deposition atmosphere on the microstructure of the films and, consequently, on their properties as capping layers.

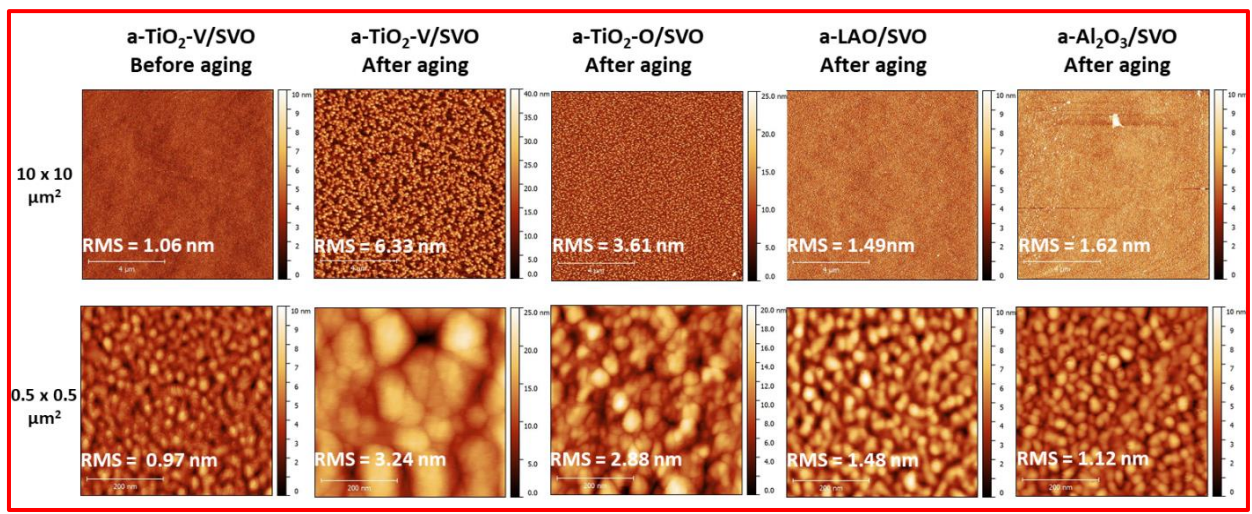

**Figure 4.** AFM images of SVO films protected with different capping layers before and after aging. The scale bar indicated in the images for the large scan is 4 µm, and 200 nm for the small scan.

For SVO films protected with LAO and $Al_2O_3$, a very smooth surface comparable to that of films after deposition was observed, with a slight increase in the roughness of the films. Moreover, the grain size is still comparable to that of the as-grown films. The negligible change in topography of films protected by LAO and $Al_2O_3$ can be attributed to their resistance to oxygen diffusion. It is probable that these two capping layers absorb oxygen until they reach stoichiometry, then turn into barriers against further oxygen diffusion.

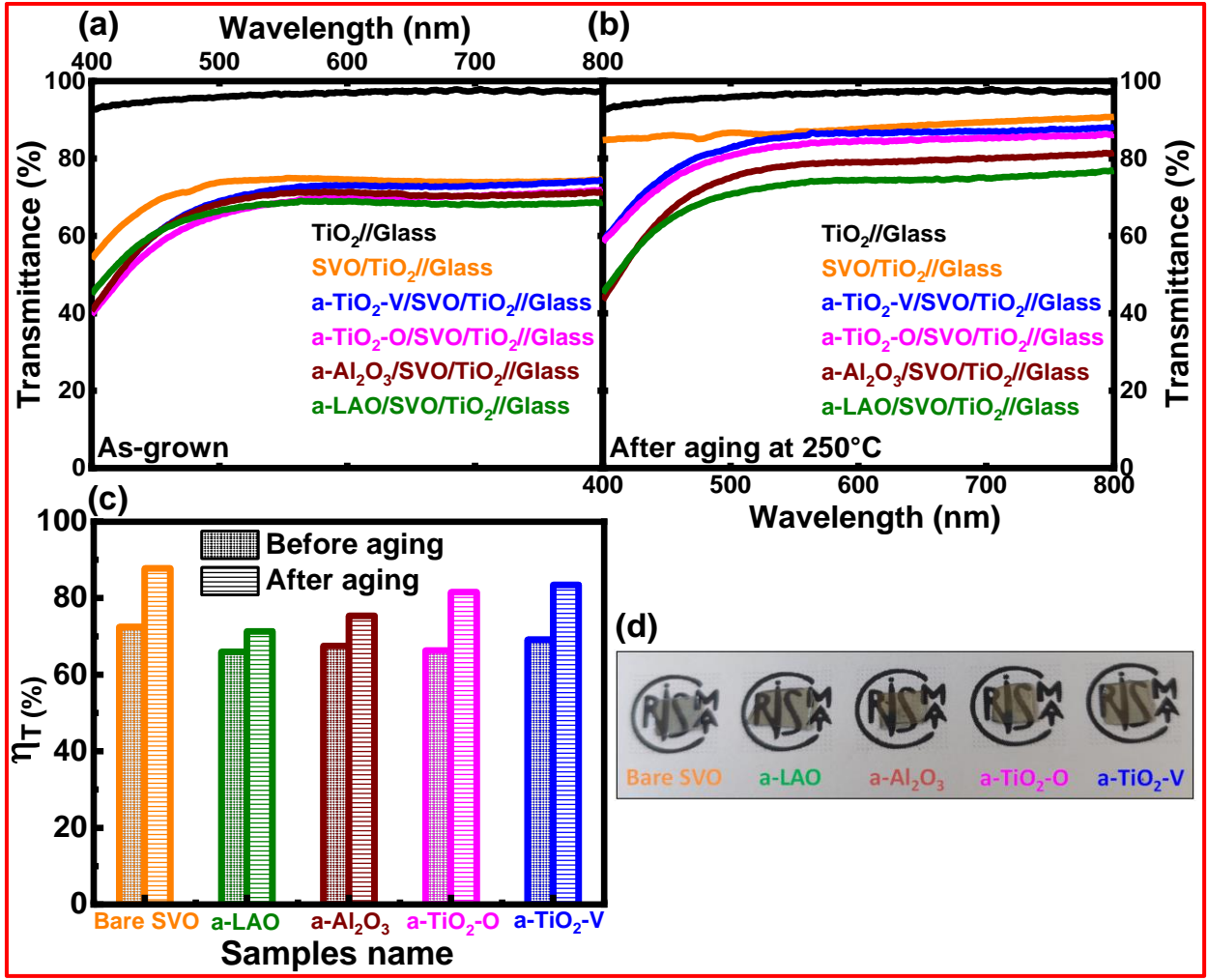

**Figure 5.** Optical transmittance spectra of polycrystalline SVO thin films with different protection layers (**a**) As-grown , (**b**) After annealing at 250 ºC for 24h under ambient atmosphere, (**c**) Variation of normalized integrated transmittance factor $\eta_T$ obtained from transmission measurement.(d) Pictures of SVO films deposited on $TiO_2$ buffered glass substrate without and with different capping layers.

Figures 5.a and 5.b show the optical transmittance spectra of SVO films without and with different capping layers, before and after annealing at 250 ºC for 24h, measured in the visible range. The corrected transmission spectra of as deposited and aged samples are plotted after eliminating the contribution from the $TiO_2$ buffer layer and the glass substrate, however, capping layers were not considered in the data correction, therefore, the transmittance of SVO films with capping layers at $\lambda$=550 nm (69, 69, 71, and 72 % for a-$TiO_2$-O, a-LAO, a-$Al_2O_3$, and a-$TiO_2$-V, respectively) is lower than that of bare SVO film (75%). Figure 5.a reveals a flat large transmittance from 800 nm down to 500 nm with a transmittance drop from 500 to 400

nm for all as deposited samples. This pronounced decrease of transmittance below 500 nm of wavelength of SVO is attributed to the presence of interband transition from Oxygen 2$p$ bands forming the valence band to the unoccupied states of the conduction band derived from the V $t_{2g}$ orbitals. It is important to note that the as-grown SVO films without and with different protection layers behave as transparent conducting materials in the visible range. Interestingly, after annealing at 250 ºC for 24h, the samples with a-LAO and a-Al$_2$O$_3$ protecting layers show a small increment of transmittance (74 and 78 % for a-LAO and a-Al$_2$O$_3$) compared to the value before aging at the same wavelength (550 nm). In contrast, the samples with TiO$_2$ protecting layers show a high value of transmittance (83 and 86 % for a-TiO$_2$-O (magenta) and a-TiO$_2$-V (blue)), which are remarkably close to that of the SVO film without protection (86%). The remarkable enhancement in the transmittance of the SVO films protected with TiO$_2$ layers is related to the formation of the over-oxidized Sr-rich V$^{5+}$ phases, i.e. Sr$_3$V$_2$O$_8$ or Sr$_2$V$_2$O$_7$, in the interface between the SVO film and the TiO$_2$ capping layers because of the oxygen diffusion. These insulating phases are known to be more transparent, as is the case with the bare annealed SVO film. The absence of charge carriers renders the material more transparent. Moreover, it should be noted that the transmittance drops at around 480 nm, related to the interband transitions from O2p to V3d in crystalline SVO films, is still detected for protected SVO films and even with TiO$_2$ capping layers while this signature has vanished for the unprotected SVO film. As the interband transitions are related to transitions in the band structure characteristic for the perovskite phase, the absence of the transmittance drop is a sign that the film has entirely lost its crystalline structure. In contrast, SVO films protected with TiO$_2$ have only a partial degradation, resulting in a SrVO$_3$ + Sr-rich V$^{5+}$ over-oxidized phase mixture, confirming again the above GIXRD, XRR and resistivity results.

Figure 5.c shows the corresponding normalized integrated transmittance factor $\eta_T$, extracted from the transmission measurement, provides further information on the variation of transmittance values before and after heat treatment in atmosphere, in the visible range. The above factor was calculated by considering the ideal transmittance of 100 % as T$_{max}$ and is given by the following formula.

$$\eta_T \ (\%) = \frac{\int_{\lambda_{min}}^{\lambda_{max}} T(\lambda) d\lambda}{\int_{\lambda_{min}}^{\lambda_{max}} T_{max}(\lambda) d\lambda} x \ 100$$

In Figure 5.c, $\eta_T$ is plotted for the different protecting layers, comparing the values before and after aging at 250 ºC for 24h. $\eta_T$ shows an increase of 8 and 10 % for the samples with a-LAO and a-Al$_2$O$_3$ protection layers, respectively. In fact, the slight variation of η$_T$ with annealing temperature can be attributed to certain oxygen exchange between the SVO film and the protection layer. PLD grown films are known to be slightly under-stoichiometric in oxygen, so that a heat treatment may allow for a healing of these defects, as was observed in monocrystalline films[11]. Anyway, the films with Al-based capping layers stay conducting, so that the annealing is even useful to achieve TCO films with outstanding functional properties. Furthermore, the η$_T$ of 82% with the electrical resistivity of ~10$^4$ $\mu\Omega$ cm is observed for SVO film with TiO$_2$ capping layers grown under oxygen atmosphere, while the highest η$_T$ value of 83% is found for the sample with the TiO$_2$ capping layer deposited in vacuum. The above value is comparable with the unprotected SVO (88%) but much higher than the samples with a-LAO (71%) and a-Al$_2$O$_3$ (75%) protection layers (as seen in Figure 5.c). The increase in transmittance value with the higher electrical resistivity (10$^4$ $\mu\Omega$.cm) indicates poor conductor/insulator like behaviour of the sample after the heat treatment.

To quantify the protection efficiency of various protecting layers used in this study, we have introduced a figure of merit (FOM) based on the η$_T$ factor and the resistivity:

$$FOM = \frac{\int_{\lambda_{min}}^{\lambda_{max}} T(\lambda)d\lambda}{\rho}$$

This factor considers the entire range of transmittance in the visible spectrum (from 400 to 800 nm) along with the resistivity ($\rho$) of the SVO films, which represents an intrinsic property of the material. Importantly, this figure of merit emphasizes sensitivity to both transparency and resistivity, assigning equal significance to both parameters. In contrast to Haacke's method[45], which typically considers only a single point in the visible spectrum (often, the transmittance at $\lambda$ = 550 nm) raised to the power of 10 and divided by the sheet resistivity, which is thickness dependent. Our figure of merit offers a more comprehensive description of the optoelectronic properties of the films compared to Haacke's method.

Figure 6.a presents the FOM for SVO films without and with different capping layers, both before and after aging. On the other hand, Figure 6.b illustrates the ratio between the FOM after aging and before aging. This comparison allows an in-depth evaluation of the efficiency of each capping layer.

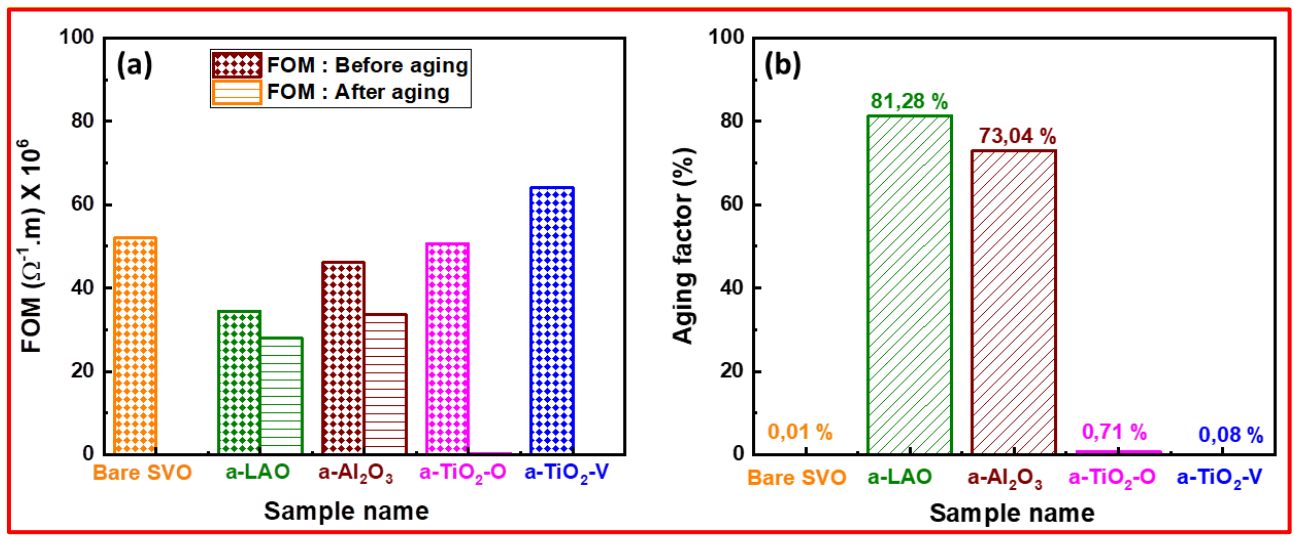

**Figure 6.** (a) Figure of merit based on the $\eta_T$ factor and the resistivity of the films. (b) The aging factor, denoted as the ratio between the FOM after aging and the FOM before aging.

Before aging, it is evident that SVO films without protection or those protected with TiO$_2$-based capping layers exhibit a higher figure of merit. This is attributed to the low resistivity and high transparency of these films. However, SVO films protected with Al-based capping layers exhibit a slightly lower figure of merit.

After aging, films protected with Al-based capping layer exhibit the highest figure of merit compared to other capping layers. Moreover, films protected with a-LAO and Al$_2$O$_3$ exhibit a decrease in figure of merit of 19% and 27%, respectively, compared to their values before aging. On the contrary, both unprotected films and those protected with TiO$_2$-based capping layers undergo a significant decline in figure of merit, approaching a reduction of 100%. The substantial degradation in their optical and electronic properties signifies that these films are no longer suitable for use as transparent conducting oxides.

Furthermore, it is essential to note that the focus lies not solely on the figure of merit's value but rather on its variation post-aging, serving as an indicator of the capping layer's efficacy. In our investigation, the observed variability underscores the superior protective performance of Al-based capping layers.

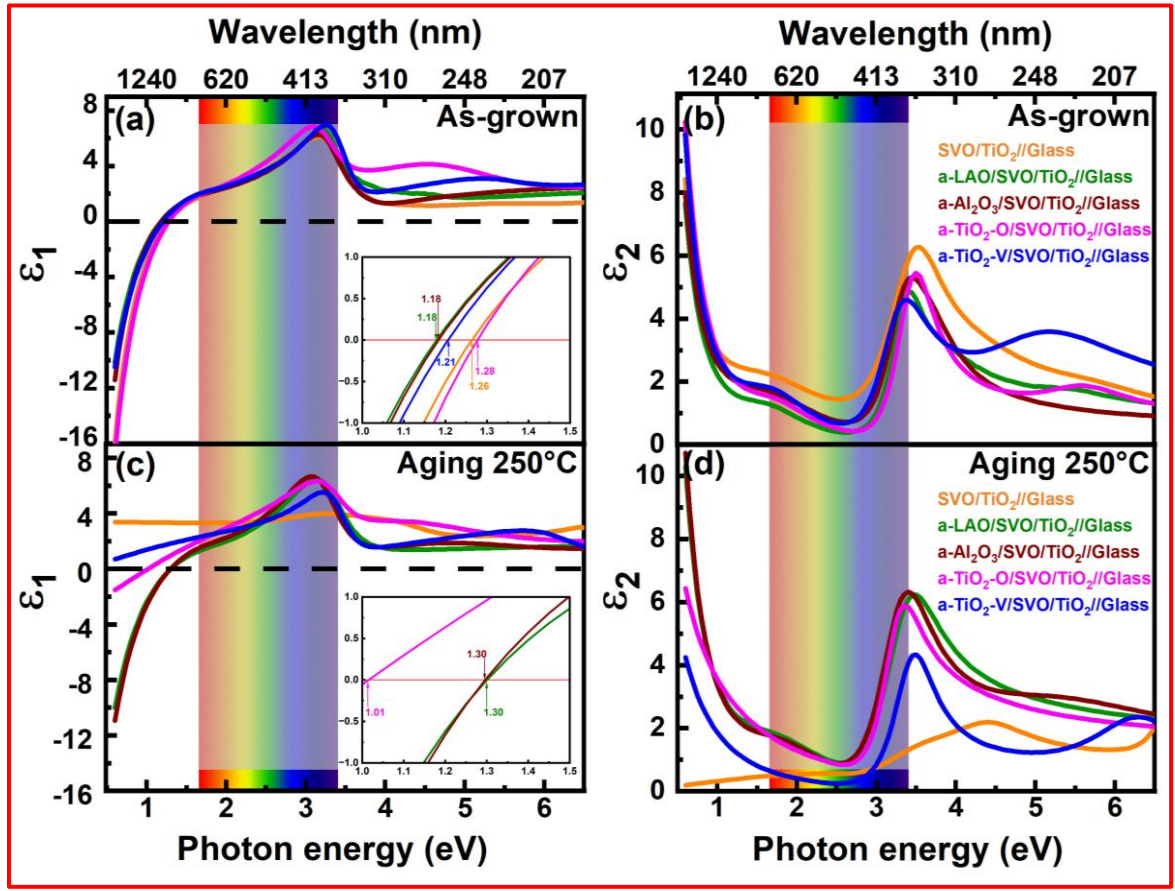

**Figure 7.** Spectroscopic ellipsometry measurements (**a, b**) Real ($\varepsilon_1$) and imaginary ($\varepsilon_2$) parts of dielectric complex function ($\varepsilon = \varepsilon_1 + i\varepsilon_2$) of as grown polycrystalline SVO thin films with different protection layers.(**c, d**) Real and imaginary parts of SVO films with various protecting layers' after annealing for 24h in air . The inset in (a and c) is a zoom between 1 and 1.5 eV.

Figures 7.a and 7.b show the real and imaginary parts of dielectric complex permittivity ($\varepsilon(\omega) = \varepsilon_1(\omega) + i\varepsilon_2(\omega)$, $\omega$ is the photon angular frequency) of as-grown polycrystalline SVO thin films with different capping layers. The $\varepsilon_1$ and $\varepsilon_2$ of the as-grown polycrystalline SVO films with and without protecting layers, share the same shapes and the same energy positions as the earlier reported results[46]. The as-grown polycrystalline SVO films with and without protection layers show a zero crossing of the real part of the dielectric constant at $\varepsilon_1(\omega) = 0$, extracted from spectroscopic ellipsometry (SE) measurements (as shown the insets in Figure 5.a). It is reported that the zero crossing of $\varepsilon_1(\omega)$ is situated at around ~ 1.3 eV for SVO on single crystals, which corresponds to the screened plasma energy ($\hbar\omega_p^*$) of the material[46,47]. The ellipsometry results show that $\hbar\omega_p^*$ of the as-grown polycrystalline SVO samples (Figure 7.a) are dependent on the type of the capping layer. The reference film, *i.e.* the as-grown unprotected SVO film, has a $\hbar\omega_p^*$ of 1.26 eV. The slightly lower $\hbar\omega_p^*$ obtained in these polycrystalline layers compared to epitaxial films indicate that the films are more transparent in the IR region, and the observed red shift is related to a certain concentration of oxygen vacancies inside the SVO films, leading to a variation of the electronic correlations. For the capping layers deposited under vacuum, the observed values are lower for all three samples (1.18 eV for both a-$Al_2O_3$ and a-LAO samples, 1.21 eV for a-$TiO_2$-V). This is coherent with the above situation, as the a-LAO, a-$Al_2O_3$ and a-$TiO_2$-V capping layers are deposited under vacuum and may attract oxygen ions from the underlying SVO film to achieve a stable stoichiometry, resulting in an even lower $\hbar\omega_p^*$. For the $TiO_2$-based capping layer grown under oxygen a-$TiO_2$-O, $\hbar\omega_p^*$ = 1.28 eV is higher than for the uncapped SVO film, because of the compensation of oxygen vacancies through the deposition atmosphere of the $TiO_2$, even at such low temperatures (100 °C). Besides the zero crossing of $\varepsilon_1(\omega)$, the enhancement of $\varepsilon_2(\omega)$ at low energies < 1 eV observed for all samples is related to the contribution of free charge carriers through the Drude peak, consistent with the conducting nature of the SVO film.

After the heat treatment at 250°C for 24h, the shape of $\varepsilon_1$ and $\varepsilon_2(\omega)$ change fundamentally. The absence of the $\varepsilon_1(\omega) = 0$ crossing and the flat $\varepsilon_2(\omega)$ evolution at low energies < 1 eV shows the absence of free charge carriers, consistent with the insulating nature of the film determined by resistivity measurements. However, the samples with a-$Al_2O_3$ (dark red) and a-LAO (dark green) protection layers in Figure 7.c exhibit zero-crossing of $\varepsilon_1(\omega)$ indicating that $\hbar\omega_p^*$ increases to reach the bulk value of 1.3 eV. This change confirms the healing of the oxygen vacancies during the heat treatment, already inferred from the slightly enhanced transmittance values in these conducting films. The aging process reduces the number of oxygen vacancies in the SVO film protected with a-LAO and a-$Al_2O_3$ and eventually resulting in stoichiometric films. Furthermore, the enhancement of $\varepsilon_2$ for a-$Al_2O_3$ and a-LAO protected SVO films in the Drude region (*i.e.* below 1 eV) is consistent with the presence of free charge carriers in these films (Figure 7.d). Hence, the samples with amorphous a-$Al_2O_3$ and a-LAO protection layers preserve their electrical as well as transparent properties even after heating at 250 °C for 24 h in air and will perform as a TCO for optoelectronic applications.

The sample a-$TiO_2$-O (magenta) shows a zero crossing of $\varepsilon_1(\omega)$, but the screened plasma frequency shifts to lower energy (1.01 eV), related to the partial degradation of the SVO film observed by the other characterisations. The Drude contribution of free charges drops correspondingly. The sample a-$TiO_2$-V (blue) shows an even stronger degradation, with a non-negative $\varepsilon_1(\omega)$ value of SVO and an even lower Drude tail, although the contribution does not disappear totally. Hence, after heat treatment, the sample with $TiO_2$ based capping layers behaves as poor electronic conductor, which agrees with the electrical resistivity results (Figure 3).

To take a closer look at the evolution of the electronic structure of SVO films, we calculated the optical conductivity before and after aging using the following formula:

$$Re\left(\sigma(\omega)\right) = \varepsilon_0 \omega \varepsilon_2(\omega)$$

The calculated real part of the optical conductivity is shown in Figure 8.

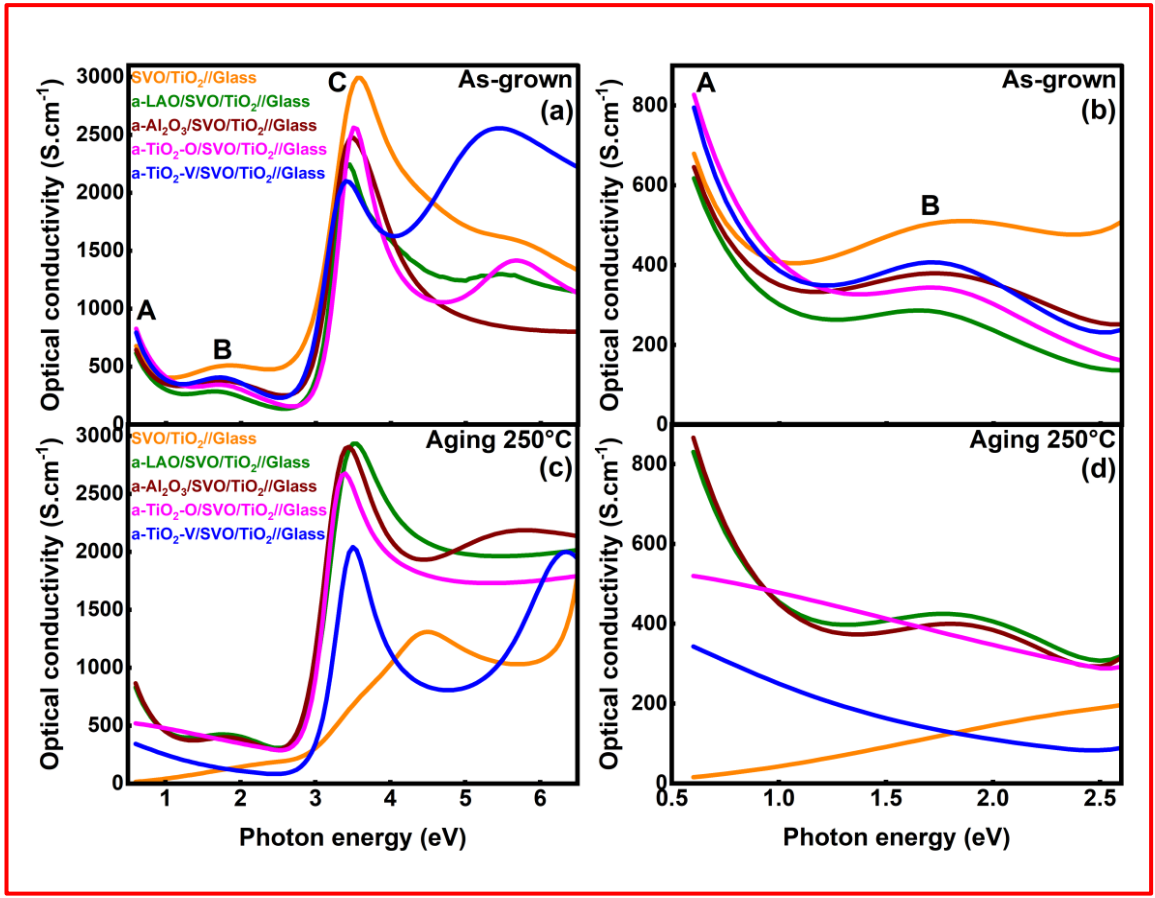

**Figure 8.** Optical conductivity of as-grown (a and b) and aged at 250°C (c and d) SVO films capped with different protecting layer.( b and d) are a zoom at A and B peaks.

In the as-grown films, the three characteristic peaks of SVO are present, with the first peak (A) corresponding to the Drude contribution. A second peak (B) appears within the range of intermediate energies around 1.7 eV and corresponds to the intraband d-d transitions from the occupied Lower-Hubbard Band (LHB) to the quasiparticle peak or to the unoccupied Upper Hubbard Band (UHB)[48]. The third peak (C) appears at higher energies (around 3.5 eV), which corresponds to the already mentioned charge transfer transitions from O2p to V3d. Regarding the optical conductivity in the Drude region (A) in the as-grown state, the SVO films protected with a-TiO$_2$ layers have the highest optical conductivity. Films protected with a-LAO and a-Al$_2$O$_3$ have the lowest optical conductivity, while the unprotected SVO film shows an intermediate value. After aging, the optical conductivity of this unprotected film approaches zero. Simultaneously, films protected with a-TiO$_2$ capping layers exhibit a significant decrease in optical conductivity. In contrast, films protected with a-LAO and a-Al$_2$O$_3$ layers exhibit an enhancement in the optical conductivity.

Moreover, we observe the suppression of the quasiparticle peak B situated at around 1.7 eV for samples without protection and with a-TiO$_2$ protection layers, which is related to the transformation of the crystalline SrVO$_3$ phase to the Sr-rich V$^{5+}$ over-oxidized phases. On the other hand, the quasiparticle peak of SVO films protected with a-LAO and a-Al$_2$O$_3$ is still observed, and its intensity is slightly enhanced which confirms the stability of SVO films when protected with Al-based protecting layers. These findings align with the observed changes in resistivity over time during the aging process.

**CONCLUSIONS**

This study has focused on the development of efficient capping layers to protect SrVO$_3$ thin films from deterioration in ambient air. A comparative study of various amorphous capping layers, including TiO$_2$ grown in different atmospheres, LAO and Al$_2$O$_3$, has been carried out. Amorphous lanthanum aluminate (a-LAO) and aluminum oxide (a-Al$_2$O$_3$) stand out as the most effective of these capping layers due to their ability to successfully preserve the original state of SVO films without affecting their structural, chemical, electrical, or optical properties. Amorphous TiO$_2$ capping layers proposed earlier[39] have shown extremely limited effectiveness compared to the Al-based capping layers, probably because of their limited ability to avoid oxygen diffusion, which leads to the chemical transformation of the conducting SrVO$_3$ phase to over-oxidized V$^{5+}$ insulating phases. This study not only contributes to the advancement of thin films protection strategies for industrial applications, but also paves the way for enhanced stability and extended lifespan of SVO-based TCOs devices in real conditions.

**METHODS**

**Samples preparation:** Polycrystalline targets of $Sr_2V_2O_7$, $TiO_2$, $Al_2O_3$ and $LaAlO_3$ were prepared by standard solid-state reaction route. A series of polycrystalline SVO films is grown with different capping layers such as amorphous $Al_2O_3$ (a-$Al_2O_3$), $LaAlO_3$ (a-LAO), $TiO_2$ under vacuum (a-$TiO_2$-V) and $TiO_2$ under oxygen partial pressure (a-$TiO_2$-O). Eagle XG slim glass substrates with 7 nm thick $TiO_2$ layer as a seed layer was used to grow *in-situ* polycrystalline SVO films. The deposition was carried out in a Pulsed Laser Deposition (PLD) setup using an ultraviolet KrF excimer laser with a wavelength of 248 nm. All SVO films, $TiO_2$ seed layers and different capping layers are deposited with a laser energy of 200 mJ corresponding to a fluence of around 1.6-2.0 J.cm$^{-2}$ and with a laser repetition rate of 5 Hz as discussed in our previous reports[21]. An optimized growth temperature of 600 °C was used for the $TiO_2$ seed layer and the SVO films while the capping layers were deposited at 100 °C to obtain amorphous layers. The $TiO_2$ seed layers and the SVO films were deposited under vacuum, with a pressure of around 2.0 x 10$^{-6}$ mbar, when the sample is at deposition temperature. The capping layers were deposited under the same atmosphere, except the a-$TiO_2$-O capping layer, where an oxygen $O_2$ partial pressure ($P_{O2}$=1x10$^{-2}$ mbar) was used. The thicknesses of the seed and capping layers were approximately 7-10 nm (2500 pulses) and 25-30 nm (5000 pulses) for SVO films, as determined by X-Ray Reflectivity (XRR) measurements.

**Structural Characterization:** The crystal structure of polycrystalline films with and without protection layers, before and after heat treatment, was studied by Grazing Incidence X-Ray Diffraction (GIXRD) using a PANalytical X'Pert Materials Research Diffractometer (MRD) with monochromatic Cu $K_{\alpha1}$ radiation and wavelength of 1.5406 Å. X-ray reflectivity (XRR) using the same instrument was used to determine thicknesses of the samples. The exact values of the film were determined by considering the positions of the maxima of the XRR oscillations, and a fit of the sinus of these positions vs the square of the order number. The slope of the fit is used to calculate the film thickness.

**Electrical characterization:** The thermal annealing of as grown samples with and without protection layers was carried out on a hot plate under ambient atmosphere, for 24 h. The samples were placed on the hot plate at room temperature, and the four-point probe configuration for the resistivity measurement was positioned on the surface of the samples. Then, to study the aging effect of these samples, the temperature of the hot plate was increased slowly until it was reached to 250 °C. By applying a current, the electrical resistance with time was recorded using a Keithley 2450 sourcemeter, once the desired temperature was reached. The initial resistance value at t = 0 min corresponds to the room temperature resistance of the sample before the beginning of the heat treatment. Finally, the electrical resistivity of the aged samples with time was calculated from the resistance values, by considering the geometrical configuration of the sample and the measurement setup.

**Optical Characterizations:** The optical transmission of the samples was measured using conventional transmission mode of a Lambda 1050 Perkin-Elmer spectrophotometer in the UV-Visible-Near Infrared (UV-Vis-NIR) range. Variable-angle spectroscopic ellipsometry (VASE) was carried out at three different incident angles 60°, 70° and 75°, in reflection mode from the photon energies from 0.6 to 6.5 eV (UV-Vis-NIR) using a HORIBA Jobin Yvon instrument. The optimized incident angle 70° from the sample normal was used to calculate the ellipsometry parameters $(\Delta, \Psi)$ and followed by the other optical parameters like the optical coefficient $(n + ik)$ and complex dielectric constant, $\varepsilon(\omega) = \varepsilon_1(\omega) + i\varepsilon_2(\omega)$ were extracted of the samples using the DeltaPsi2 software. The collected SE data in the near IR to UV range was fitted with the configuration of air/capping layer/SVO film/substrate, where the Eagle glass substrate with buffered $TiO_2$ layer was separately fitted. It is important to note that the thicknesses of each layer such as $TiO_2$ buffer layer, SVO film and all capping layers were determined using XRR measurements and their values were inserted in the model during the fitting. New amorphous oscillator functions were used to model the insulating capping layers and $TiO_2$ buffer layer on the glass substrate. Besides, a Drude oscillator and an extra Lorentz oscillator with new amorphous oscillators were used for the metallic SVO thin film. A constant value $\varepsilon_\infty$ is added to the real part $\varepsilon_1(\omega)$ to fit the model further at even higher energy.

## ACKNOWLEDGEMENTS

The authors would like to acknowledge financial support from the CNRS prématuration project CoCOT, and the Normandie region through the projects RIN PLDsurf and Cibox, and the funding of the PhD thesis of M.M.